# Analysis of Longitudinal Data with Missing Values in the Response and Covariates Using the Stochastic EM Algorithm

Ahmed M. Gad[a] and Nesma M. Darwish[b]

[a] *Business Administration Department, Faculty of Business Administration, Economics and Political Science, The British University in Egypt, Cairo, Egypt*

[b] *Management and information system Department, Higher Institute of computer and information technology, El_Shrouk Academy, Cairo, Egypt*

Corresponding Author: ahmed.gad@feps.edu.eg (Ahmed Gad)

**Abstract**

In longitudinal data a response variable is measured over time, or under different conditions, for a cohort of individuals. In many situations all intended measurements are not available which results in missing values. If the missing value is never followed by an observed measurement, this leads to dropout pattern. The missing values could be in the response variable, the covariates or in both. The missingness mechanism is termed non-random when the probability of missingness depends on the missing value and may be on the observed values. In this case the missing values should be considered in the analysis to avoid any potential bias. The aim of this article is to employ multiple imputations (MI) to handle missing values in covariates using. The selection model is used to model longitudinal data in the presence of non-random dropout. The stochastic EM algorithm (SEM) is developed to obtain the model parameter estimates in addition to the estimates of the dropout model. The SEM algorithm does not provide standard errors of the estimates. We developed a Monte Carlo method to obtain the standard errors. The proposed approach performance is evaluated through a simulation study. Also, the proposed approach is applied to a real data set.

# 1. Introduction

Longitudinal studies are common in many disciplines including bioscience field. In longitudinal studies a response variable is measured repeatedly for every individual. In these cases, one variable is the underlying characteristic or measurement. Longitudinal studies have many advantages and disadvantages. The main advantage of longitudinal studies is that it can distinguish changes over time within individuals and enabling direct study of that change. The main disadvantage of longitudinal studies is early withdrawal of some individuals. This result in dropout missing data. The missing values are very common in longitudinal data. The missing values may occur in the response variable or in the covariates or both. It is common in practice to have missing values in the response variable and in the covariates. Little (1995) reviews some approaches where the covariates are completely observed. In this article we focus on missingness in response and covariates.

The selection model is used to model longitudinal data in presence of missing values. This model factorizes the joint distribution of the response $Y_i$ and the missing data mechanism indicator $R_i$ to product of the marginal distribution of $Y_i$ and conditional distribution of $R_i$ given $Y_i$. Thus

$$f(Y_i, R_i|\theta, \Psi) = f(Y_i|\theta)P(R_i = r_i|Y_i, \Psi),$$

where $\theta$ is a vector containing the model parameters, $P(R_i = r_i|Y_i, \Psi)$ is the distribution that characterizes the missing data mechanism, and $\Psi$ is a vector of parameters that govern the missing data mechanism. According to Rubin (1976) and Little and Rubin (1987) the missing data mechanism is missing completely at random (MCAR) if $R_i$ and $Y_i$ are independent,

$$P(R_i = r_i|Y_{i,obs}, Y_{i,mis}\Psi) = P(R_i = r_i|\Psi) \ ;$$

the missing data mechanism is missing at random (MAR) if the conditional distribution of $R_i$ given $Y_i$ depends only on the observed, $Y_{i,obs}$ i.e.

$$P(R_i = r_i|Y_{i,obs}, Y_{i,mis}\Psi) = P(R_i = r_i|Y_{i,obs}, \Psi) \ ;$$

and nonrandom (informative) otherwise.

In dropout pattern, Diggle and Kenward (1994) propose a selection model for longitudinal data with nonrandom dropout. They fit a normal linear model for the response variable, $Y_i$, and a logistic model for the probability of dropout. The probability of dropout at time $d_i$ is modelled as a function of the response at time $d_i$ and the observed measurements (history); that is,

$$P(D_i = d_i|history) = P_{d_i}(H_{d_i}, y_{d_i}, \Psi).$$

Also, they suggest using the logistic model for the dropout process as

$$logit\{P_{d_i}\{H_{d_i}, y_{d_i}, \Psi\}\} = \Psi_0 + \sum_{j=1}^{d_i} \Psi_j \ y_{d_i-j+1}$$

The stochastics EM algorithm (SEM) has been proposed by Celuex and Diebolt (1985) to overcomes the main difficulty of the EM algorithm by avoiding explicit calculation of the E-step. The SEM algorithm iterates two steps: the S-step and the M-step. In the S-step the missing values are imputed using a single draw from the conditional distribution of the missing data given the observed data. In the M-step, the log-likelihood function of the pseudo-complete can be maximized using a standard maximization procedure. The process is repeated for sufficient number of iterations. The SEM algorithm has many advantages. These include the ability to deal with multimodality of the likelihood surface (Ip, 1994). Also, the sequence of the parameter estimates forms a Markov chain which converges reasonably quickly to its unique stationary distribution (Diebolt and Ip, 1996). The parameter estimates $\tilde{\beta}$ can be obtained as the average of the sequence of each parameter, excluding the early iterations as burn-in period (Diebolt and Ip,1996). Gad and Ahmed (2006) apply the SEM algorithm to longitudinal data with dropout in the response.

The SEM algorithm does not provide the standard errors of the parameter estimates. In literature there are many methods to overcome this drawback. Louis (1982) propose a formula relates the observed information matrix to the conditional expectation of the second derivatives of complete data log-likelihood function and the covariance of the first derivatives of complete data log-likelihood function. The use of this formula requires evaluating integrals, which is not easy task. Efron (1994) suggest using simulation (the Monte Carlo method) to approximate the integrations. The missing values are simulated from their conditional distribution and then each integration is evaluated by its empirical version.

Erler et al. (2016) evaluate the performance of multiple imputation chained equation using different strategies to include a longitudinal outcome into the imputation models and compare it with a fully Bayesian approach. Nooraee et al. (2018) investigate a hybrid approach which is a combination of maximum likelihood and multiple imputation, i.e. scales from the imputed data are eliminated if all underlying items were originally missing. Abdelwahab et al. (2019) propose a sensitivity analysis index for shared parameter models in longitudinal studies. Darwish et al. (2020) propose using multiple imputation for missing at random cross-sectional covariates. They employ a shared parameter model to fit the response variable in the presence of non-random dropout.

The aim of this article is to suggest a multiple imputation approach for cross-sectional covariates. The selection model is adopted for fitting linear regression model between longitudinal response and cross-sectional covariates where both the response and the covariates have missingness. The rest of the article is organized as follows. In Section 2 we present the two common multiple imputation methods that can be used to handle missingness in covariates. In Section 3 the proposed approach is described in addition to the Monte Carlo method as a way for obtaining the standard error estimates. In section 4, a simulation study is presented to validate the proposed approach. In section 5 the proposed approach is applied to a real data. Finally, Section 6 presents the conclusion and future work.

## 2. Multiple Imputations (MI) Methods

Grannell and Murphy (2011) discussed the application of the MI using four different methods which are applied practically using the SOLAS package. Salfran and Spiess (2015) described some of the most common imputation methods included in software packages. We provide more details about used imputation methods

## 2.1 Regression-based imputation

In the regression method, a regression model is fitted for each variable with missing values. Based on the resulting model, a new regression model is then drawn and is used to impute the missing values for the variable since the data set has a monotone missing data pattern, the process is repeated sequentially for variables with missing values. That is, for a variable $Y_j$ with missing values, a model

$$Y_j = \beta_0 + \beta_1 X_1 + \ldots + \beta_k X_k \tag{1}$$

is fitted using observations with observed values for the variable $Y_j$ and its covariates $(X_1,\ldots,X_k)$. The fitted model includes the regression parameter estimates as follow,

$$\hat{\beta} = (\hat{\beta_0}, \hat{\beta_1},\ldots,\hat{\beta_k}) \text{ and the association covariance matrix } \hat{\sigma}^2, V_j. \tag{2}$$

The following steps are used to generate imputed values for each imputation:

1- New parameter $\beta_* = (\beta_{0*}, \beta_{1*},\ldots,\beta_{k*})$ and $\sigma^2_{*j}$ are drawn from the posterior predictive distribution of the parameters. That is, they are simulated from $\hat{\beta} = (\hat{\beta_0}, \hat{\beta_1},\ldots,\hat{\beta_k})$ and the association covariance matrix $\hat{\sigma}^2, V_j$. The variance is drawn as

$$\sigma^2_{*j} = \hat{\sigma}^2 (n_j - k - 1)/g, \tag{3}$$

where $g$ is $\chi^2_{n_j-k-1}$ random variate and $n_j$ is the number of non-missing observation for $Y_j$. The regression coefficients are drawn as

$$\beta_* = \hat{\beta} + \sigma^2_{*j} V'_{hj} Z, \tag{4}$$

where $V'_{hj}$ is the upper triangular matrix in the Cholesky decomposition of $V_j = (V'_{hj} V_{hj})$ and $Z$ is a $k+1$ independent random normal variates.

2- The missing values are then replaced by

$$y_{i*} = \beta_{0*} + \beta_{1*} x_1 + \beta_{2*} x_2 + \ldots + \beta_{k*} x_k + z_i \sigma_{*j}, \tag{5}$$

where $(x_1, x_2,\ldots,x_k)$ are the values of the covariates and $z_i$ is a simulated normal deviate.

The regression method, is especially simple to extend for longitudinal data with dropout (or withdrawal or attrition), thus generating a monotone pattern of amusingness. A sequential imputation procedure can be used, starting with the first measurement (assumed complete) in which, at each time point, previously imputed values (possibly a subset of these) are used in the imputation model as predictors for future values. The sequential imputation method works as follows, assuming that the data are complete at the first time point the simple imputation method summarized above can be used to impute the data missing at the second time of measurement. This process is repeated at the third time point and sequentially up to the final time point. For each time of measurement, previously imputed values are introduced as explanatory variables into the regression model. One pass through the sequence generates a single set of imputations, and the whole process is repeated $M$ times to obtain $M$ completed data sets. In principle, the normal-based regression model can be replaced by appropriate models for other types of outcome, for example logistic regressions for binary data, or proportional odds models for ordinal data.

## 2.2 Predictive Mean Matching method (PMM)

The predictive mean matching method can also be used for imputation. It is like the regression method except that for each missing value, it imputes an observed value which is closest to the predicted value from the simulated regression model (Rubin 1987). The predictive mean matching method ensures that imputed values are plausible and may be more appropriate than the regression method if the normality assumption is violated. Following the description of the predictive mean matching method, the steps are used to generate imputed values as the regression method, but it need set of $k_0$ observations whose corresponding predicted values are closest to $y_{i*}$ is generated. The missing value is then replaced by a value drawn randomly from these $k_0$ observed values.

## 3. The Proposed Approach

First we handle missingness in covariates through multiple imputation using the regression method or predictive mean matching method.

- **Imputing continuous cross-sectional covariates with monotone missingness using regression method.**

Depending on complete out-come from previous models we impute missingness in cross-sectional covariates as follows,

$p(x_{i,mis} \mid y_i, x_{i,obs})$

$x_i = \beta_0 + \beta_1 y_i + \beta_{2i} x_{i,obs}$

$\hat{\beta} = (\hat{\beta_0}, \hat{\beta_1}, \hat{\beta_{2i}})$ and the association covariance matrix $\hat{\sigma}^2, V_i$.

The following steps are used to generate imputed values for each imputation:

1. New parameter $\beta_* = (\beta_{0*}, \beta_{1*}, \beta_{2i*})$ and $\sigma^2_{*i}$ are drawn from the posterior predictive distribution of the parameters. That is, they are simulated from $\hat{\beta} = (\hat{\beta_0}, \hat{\beta_1}, \hat{\beta_{2i}})$ and the association covariance matrix $\hat{\sigma}^2, V_i$. The variance is drawn as

$$\sigma^2_{*i} = \hat{\sigma}^2 (n_i - k - 1)/g,$$

where g is $\chi^2_{n_j - k - 1}$ random variate and $n_i$ is the number of non-missing observation for $X_i$. The regression coefficients are drawn as

$$\beta_* = \hat{\beta} + \sigma^2_{*i} V_{hj}' Z$$

where $V_{hj}'$ is the upper triangular matrix in the Cholesky decomposition of $V_j = (V_{hj}'V_{hj})$ and $Z$ is a k+1 independent random normal variates.

2. The missing values are then replaced by

$$x_{i*} = \beta_{0*} + \beta_{1*}y_i + \beta_{2*}x_{i,obs} + Z_i\sigma_{*i}^2,$$

where $y_i$ longitudinal response and $x_{i,obs}$ another complete covariates are consider the values of the covariates and $z_i$ is a simulated normal deviate.

- **Imputing continuous cross-sectional covariates with monotone missingness using predictive mean matching method.**

Following the description of the model in the previous section Monotone Regression Methods, all the previous steps are used to generate imputed values with extra steps,

1- A set of $k_0$ observations whose corresponding predicted values are closest to $x_{i*}$ is generated.
2- The missing value is then replaced by a value drawn randomly from these $k_0$ observed values.

Second the SEM algorithm can be applied using the pseudo complete covariates using the two steps; S-step and M-step.

**The S-Step:**

In this step, the missing data are simulated from their conditional distribution, given the observed values and the current parameter estimates, $Y_{i,mis} \sim f(Y_{i,mis}|Y_{i,obs}, R_i; \theta)$. This distribution does not have a standard form, hence it is not possible to use the direct simulation. To overcome this problem, an accept-reject procedure is developed to generate the missing values $y_{i,mis}$: This procedure mimics the dropout process assuming that the postulated dropout model is correct. A draw from $f(y_{i,mis}|Y_{i,obs}, \theta^{(t)})$ is obtained instead of $f(y_{i,mis}|Y_{i,obs}, R_i, \theta^{(t)})$ and then using the proposed procedure, this value can be accepted or rejected (Metropolis Hasting procedure (Gilks et al, 1996)). The steps of this procedure can be summarized as follows:

1. Generate a candidate value, $y^*$; from the conditional distribution function $f(y_{i,mis}|Y_{i,obs}, \theta^{(t)})$ which is normal distribution. Only the first dropout is simulated and the remaining dropout values are considered missing at random (Gad, 1999).
2. Calculate the probability of dropout for the candidate value $y^*$, according to the dropout model, where

$P(D_i = d_i|H_id_i) = \Psi_0 + \Psi_1 y_{d_i} + \sum_{j=2}^{d_i} \Psi_j y_{i,d_i+1-j},$

where the parameters $\Psi_j$ are fixed at the current values $\Psi_j^{(t)}$. Let us denote this probability of dropout, $P(D_i = d_i|H_id_i)$, as $P_i$.
3. Simulate a random variable U from the uniform distribution on the interval [0,1],

$$U \sim U[0,1],$$

then take $y_{i,mis} = y^*$ if $U \leq P_i$; otherwise repeat step1.

**The M-Step**

It consists of two sub-steps; the logistic step (M1-step) and the normal step (M2-step).

- In the logistic step (M1-step), the MLE.s for the dropout logistic model

$$\text{logit}\{P_i\} = \Psi_0 + \Psi_1 y_{d_i} + \sum_{j=2}^{d_i} \Psi_j\, y_{i,d_i+1-j}$$

 are obtained. The iteratively reweighted least squares method for finding the MLE of binary data models (McCullagh and Nelder,1989) can be used.

- In the M2 step, MLEs for the model parameters can be obtained using an appropriate optimization approach for incomplete data such as Newton- Raphson, Scoring method and Jennrich and Schluchter algorithm (1986). Newton- Raphson method is used in this article. The obtained estimates are the average as we have M imputed covariates,

$$\overline{\hat{\beta}_I} = \frac{1}{10}\sum_{m=1}^{10} \hat{\beta}_I$$

**Standard errors**

The Monte Carlo method is proposed and developed for evaluating the standard errors of the SEM estimates. The information matrix, according to Louis' formula, can be approximated by

$$I(\theta) = E\left(-\frac{\partial^2 l(\theta|Y_{obs}, Y_{mis})}{\partial\theta\partial\theta}\bigg|Y_{obs}\right) - cov\left(\frac{\partial l(\theta|Y_{obs}, Y_{mis})}{\partial\theta}\bigg|Y_{obs}\right)$$

$$= -E - C,$$

where $\boldsymbol{\theta}$ is fixed at the stochastic EM estimates and $\boldsymbol{l(\theta|Y_{obs}, Y_{mis})}$ is the log-likelihood function.

Efron (1994) suggest using simulation (the Monte Carlo method) to approximate the integrations in above equation. The missing values are simulated from their conditional distribution and then integrations are evaluated by their empirical versions. We adopt the same idea by simulating *M* identically distributed samples, $q_1, q_2, ...., q_M$ from the conditional distribution of the missing values given the observed values and the parameters estimates, *f(Y_{imis}|Y_{iobs},R_i)*. Louis' formula can be approximated by its empirical version, i.e.

$$E \sim \frac{1}{M}\sum_{j=1}^{M} \frac{\partial^2 l(\theta|Y_{obs}, R, q_j)}{\partial\theta\partial\theta}$$

and

$$C \sim cov\left(\frac{\partial l(\theta|Y_{obs}, R, q_j)}{\partial \theta}\right),$$

where the parameters $\theta = (\beta, \alpha, \psi)$ is fixed at the SEM estimates, $\hat{\theta} = (\hat{\beta}, \hat{\alpha}, \hat{\psi})$.

Having the *M* pseudo-complete data, the first and second order derivatives of the log-likelihood function are evaluated for each sample, and then it is possible to calculate the quantities *E* and *C* and hence the information matrix. The inverse of the information matrix is the covariance matrix of the stochastic EM estimates. The standard error estimates are the square root of the main diagonal elements of this matrix.

## 4. Simulation Study

The aim of this simulation is to investigate the performance of the proposed approach. A complete longitudinal outcome $y_{ij}$, for the subject *i* at the time point *j*, is generated from the following model $y_{ij} = \beta_0 + \beta_1 x_{ij} + \varepsilon_{ij}$, where *i=1,2,…,n* and *j=1,2,…,t*. The continuous cross-sectional covariates $x_{ij}$ are generated from the standard normal distribution. These covariates are independent from the error terms $\varepsilon_{ij}$. The error terms $\varepsilon_{ij}$ are assumed to follow a normal distribution with a mean of zero and $\sigma_e^2 = 0.5$. The number of subjects (the sample size) is fixed at 25 and 50 subjects. The time points are restricted at 5. The parameters are fixed at $\beta_0 = 5$ and $\beta_1 = 10$. The simulation is replicated 2000 replications. The missing values in the cross-sectional covariates are generated according to the logit model:
$$\text{logit}(X_i) = \eta_0 + \eta_1 X_{i-1}.$$
The parameters are fixed at $\psi = (-5, .06,)$. The missing values in the response are generated according to the model:
$$\text{logit}(r_{ij} = 1|\Psi) = \Psi_0 + \Psi_1 Y_{ij-1} + \Psi_2 Y_{ij-2}.$$
The parameters are assumed to be $\Psi = (-17, 0.11, 0.13)$. All subjects are assumed to be observed at the first time point $j = 1$. The covariance structure, of the response, is assumed to be autoregressive of order 1 with $\rho$=0.7 and $\sigma$ =6.0.

We apply the proposed approach where multiple imputation to cross-sectional covariates with number of imputations *M=10*. The final parameter estimates are obtained as the average over the multiply imputed data sets, i.e.

$$\bar{\hat{\beta}}_I = \frac{1}{10}\sum_{m=1}^{10} \hat{\beta}_I.$$

Table 1 and Table 2 present the results assuming the covariates are complete, and the sample size is 25 and 50, respectively. Table 3 and Table 4 present the results assuming that there are missing values in the covariates, and the sample size is 25 and 50, respectively.

*Table 1: Parameter estimates (Est) and the relative bias (RB); sample size n= 25, complete covariates*

|  | $\beta_0$ | $\beta_1$ | $\rho$ | $\sigma$ | $\Psi_0$ | $\Psi_1$ | $\Psi_2$ |
|---|---|---|---|---|---|---|---|
| True parameter | 5 | 10 | 0.7 | 6 | -17 | 0.11 | 0.13 |
| Est. | 5.13 | 9.61 | 0.64 | 5.8 | -15.02 | 0.12 | 0.11 |
| RB | 0.026 | 0.039 | 0.085 | 0.033 | 0.116 | 0.090 | 0.153 |

*Table 2: Parameter estimates (Est) and the relative bias (RB); sample size n= 50, complete covariates*

|  | $\beta_0$ | $\beta_1$ | $\rho$ | $\sigma$ | $\Psi_0$ | $\Psi_1$ | $\Psi_2$ |
|---|---|---|---|---|---|---|---|
| True parameter | 5 | 10 | 0.7 | 6 | --17 | 011 | 013 |
| Estimate | 5.34 | 9.67 | 0.63 | 5.81 | -16.1 | 0.12 | 0.12 |
| Relative Bias | 0.068 | 0.033 | 0.1 | 0.031 | 0.052 | 0.090 | 0.076 |

*Table 3: Parameter estimates (Est) and the relative bias (RB); sample size n= 25, MAR covariates*

|  | $\beta_0$ | $\beta_1$ | $\rho$ | $\sigma$ | $\Psi_0$ | $\Psi_1$ | $\Psi_2$ |
|---|---|---|---|---|---|---|---|
| True parameter | 5 | 10 | 0.7 | 6 | --17 | 011 | 013 |
| PPM | | | | | | | |
| Est | 4.54 | 10.09 | 0.65 | 5.75 | -15.45 | 0.13 | 0.10 |
| RB | 0.092 | 0.009 | 0.071 | 0.041 | 0.091 | 0.181 | 0.230 |
| Regression | | | | | | | |
| Est | 4.80 | 9.76 | 0.62 | 5.85 | -14.55 | 0.10 | 0.11 |
| RB | 0.04 | 0.024 | 0.114 | 0.025 | 0.144 | 0.090 | 0.153 |

*Table 4: Parameter estimates (Est) and the relative bias (RB); sample size n= 50, MAR covariates*

|  | $\beta_0$ | $\beta_1$ | $\rho$ | $\sigma$ | $\Psi_0$ | $\Psi_1$ | $\Psi_2$ |
|---|---|---|---|---|---|---|---|
| True parameter | 5 | 10 | 0.7 | 6 | --17 | 011 | 013 |
| PPM | | | | | | | |
| Est | 5.03 | 10.06 | 0.67 | 5.81 | -16.01 | 0.112 | 0.12 |
| RB | 0.006 | 0.006 | 0.042 | 0.03 | 0.058 | 0.018 | 0.076 |
| Regression | | | | | | | |
| Est. | 4.97 | 10.006 | 0.69 | 5.86 | -15.55 | 0.091 | 0.099 |
| RB | 0.006 | 0.0006 | 0.0142 | 0.02 | 0.085 | 0.172 | 0.238 |

From Tables 1 and 2 we can see that the parameter estimates, for both sample sizes, perform well in terms of the relative bias. Tables 3 and 4 show similar results for both sample sizes. As a conclusion depending on this simulation results, the SEM estimates with multiple imputations to MAR in covariates using the regression method has the best performance in terms relative bias. Hence the performance of the proposed approach is acceptable even in relatively small sample sizes.

## 5. **Application: Interstitial Cystitis Data Base (ICDB)**

The Interstitial Cystitis Data Base (ICDB) have been used by Yang and Kang (2010). Propert et al. (2000) described the main characteristics of the data. 637 patients are enrolled in the study. The study covers the period from January 1993 to November 1997. Patients are followed for symptoms of pain, urgency, and urinary frequency during that period. Patients are asked to rate each variable, in the last week, on a scale from 0 to 9. The value 0 means lowest severity and 9 means the maximum severity. In addition, the patients are required to rate the same variables in three consecutive days. The averages of the study variables over the three days are also recorded. The aim of the study is to explore the effect of continuous covariates on the continuous response (urgency).

Preliminary analysis of these data (Yang and Kang, 2010) depends on the observations collected in the first 36 months of the study. Yang and Kang (2010) treat the urgency frequency as a continuous variable whereas urinary frequency treated as a discrete variable.

There are missing values in the response variable (urgency). There are both dropout pattern and intermittent pattern. In this article we focus on dropout pattern, so we omit the intermittently missing values. This is result in a reduced sample of 450 patients. All continuous covariates are complete, and we generated missing values in the covariates. A brief description of the covariates is given in the Table 5.

*Table 5: The definition of the continuous covariates used in ICDB data.*

| Variable | Definition |
|---|---|
| Age | Patient age |
| UROD_7 | Volume at first sensation |
| UROD_9 | Volume at maximal capacity |

Figure 1 presents a histogram of the continuous outcome plotted and compared with normal density function with mean 4.25 and variance 2.13. As a pre-analysis step we checked the adequacy of the model to fit the data.

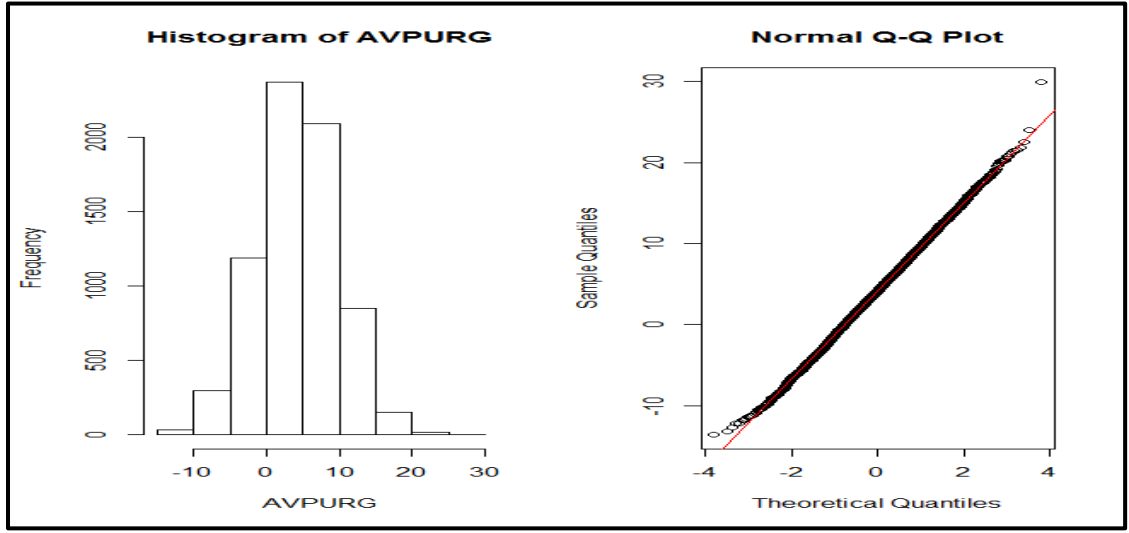
*Figure 1: A histogram of the continuous outcome with normal and normal Q-Q plot.*

We adopt a model that allows each covariate to have its own effect. That is
$$Y_i = \beta_0 + \beta_1 x_1 + \beta_2 x_2 + \beta_3 x_3 + \varepsilon_{ij}.$$
The AR(1) covariance structure is used, the elements of the covariance matrix, $\sigma_{ij} = \sigma^2 \rho^{|i-j|}$.

For the missing data mechanism, we use the linear logistic regression model. To keep the model simple only the previous and the current outcomes are included, that is
$$\text{logit}(r_{ij} = 1|\Psi) = \Psi_0 + \Psi_1 Y_{ij-1} + \Psi_2 Y_{ij-2}.$$
The proposed approach is applied to data. The results are given in Tables 6 and 7.

*Table 6: The SEM estimates and Standard errors (SE) for the urgency response without imputation to MAR covariates*

|  | Estimate | SE | p-value |
|---|---|---|---|
| Intercept ($\beta_0$) | 0.3367 | 0.07 | < 0.000 |
| UROD_7($\beta_1$) | -0.001 | 0.16 | < 0.000 |
| UROD_9 ($\beta_2$) | -0.003 | 0.17 | < 0.0640 |
| Age ($\beta_3$) | 0.05 | 0.081 | < 0.004 |
| $\rho$ | 0.12 | 0.079 | < 0.000 |
| $\sigma$ | 4.59 | 0.13 | 0.030 |
| $\Psi_0$ | -1.22 | 0.068 | < 0.000 |
| $\Psi_1$ | 0.13 | 0.079 | < 0.000 |
| $\Psi_2$ | 0.18 | 0.098 | < 0.000 |

Diggle and Kenward (1994) noticed that in non-random models, dropout tends to depend on the difference between the current and previous measurements, $Y_{ij-1} - Y_{ij}$. Using this idea the estimated model for the missing data mechanism can be viewed as:

$$logit(P) = -1.22 + 0.13Y_{ij-1} + 0.18Y_{ij},$$
$$= -1.22 + 0.13(Y_{ij} - Y_{ij-1}) + 0.31Y_{ij},$$

From this model, the positive coefficient (0.13) of the difference between $Y_{ij}$ and $Y_{ij-1}$ also indicates that the response whose urgency increased are more likely to be missing.

*Table 7: The SEM estimates and Standard errors (SE) for the urgency response with MAR missingness in Covariate*

|  | Estimate | SE | p-value |
|---|---|---|---|
| **PMM** | | | |
| Intercept($\beta_0$) | 0.445 | 0.029 | < 0.000 |
| UROD_7($\beta_1$) | -0.005 | 0.01 | < 0.000 |
| UROD_9 ($\beta_2$) | -0.0078 | 0.127 | < 0.001 |
| Age ($\beta_3$) | 0.03 | 0.156 | < 0.000 |
| $\rho$ | 0.68 | 0.011 | < 0.000 |
| $\Psi_0$ | -2.8 | 0.09 | 0.000 |
| $\Psi_1$ | 0.067 | 0.068 | < 0.000 |
| $\Psi_2$ | 0.31 | 0.09 | < 0.000 |
| $\sigma$ | 5.42 | 0.12 | < 0.000 |
| **Regression** | | | |
| Intercept ($\beta_0$) | 0.5445 | 0.0231 | < 0.000 |
| UROD_7($\beta_1$) | -0.0405 | 0.0023 | < 0.000 |
| UROD_9 ($\beta_2$) | -0.0008 | 0.0001 | < 0.000 |
| Age ($\beta_3$) | 0.0222 | 0.0009 | < 0.000 |
| $\rho$ | 0.0334 | 0.0021 | < 0.000 |
| $\Psi_0$ | -1.0234 | 0.0223 | 0.000 |
| $\Psi_1$ | 0.4559 | 0.0101 | < 0.000 |
| $\Psi_2$ | 0.124 | 0.0033 | < 0.000 |
| $\sigma$ | 4.98 | 0.098 | < 0.000 |

Based in the results in Table 6 and Table 7, the positive values for the parameter $\Psi_2$ imply that high values of the urgency are more likely to be missing. We can conclude that the null-hypothesis that, $\Psi_2 = 0$ cannot accepted; this may be evidence for non-random dropout. Also $\Psi_1$ is significantly different from 0. This indicates the importance of the response at the previous time point. Also handling MAR in covariates improves results instead of ignoring them, as ignoring the MAR in covariates at Table 6, it was an insignificant effect of UROD_9 on urgency, after handling missingness with two MI methods the effect of UROD_9 on urgency become significant as in Table 7.

### 6. Conclusion and Future Work

Most literature, in longitudinal studies with missing values, focus on missingness in the longitudinal response or in the covariates. However, in practice it is possible to have missingness in the longitudinal response and in the covariates at the same time. This article proposes two methods to deal with missingness in both the longitudinal response and covariates at the same time. In this paper we proposed a selection model (Diggle and Kenward, 1994) for longitudinal data with non-ignorable missing values of the response with multiple imputation to missingness on covariates. The proposed model covers the case of the dropout missingness. The obtained likelihood function is intractable and not easy to be maximized. To overcome this difficulty, we suggest using the Stochastic EM algorithm. The proposed approach is applied to a data set from (The Interstitial Cystitis Data Base (ICDB)). The approach can be easily implemented in many fields where the missingness process is suspected to be non-ignorable. The case of intermittent pattern, which has less attention in the literature compared to the dropout, is a very challenging topic for future work.

### Acknowledgments

The authors would like to thank Professor Kathleen Joy Propert at University of Pennsylvania, School of Medicine for providing guidelines about how to obtain the ICDB data. The ICDB data reported here were supplied by the National Institute of Diabetes and Digestive and Kidney Diseases (NIDDK) Central Repositories.

### Statements and Declarations

No funding or competing interests.